\documentclass[12pt]{article}

\begin{document}

\thispagestyle{empty}

\noindent {\small CITUSC/01-001\hfill \hfill hep-th/0103042 \newline
}

{\small \hfill }

{\vskip0.5cm}

\begin{center}
{\Large \textbf{High Spin Gauge Fields and Two-Time Physics\\[0pt]
}}

\bigskip

{\vskip0.5cm}

\textbf{Itzhak Bars}$^{a}$ \textrm{and} \textbf{Cemsinan Deliduman}$^{b}$

{\vskip0.5cm}

$^{a)}$\textsl{CIT-USC Center for Theoretical Physics \& Department of
Physics}

\textsl{University of Southern California,\ Los Angeles, CA 90089-2535, USA}

\texttt{bars@physics.usc.edu}

{\vskip0.5cm}

$^{b)}$\textsl{Feza G\"{u}rsey Institute}

\textsl{\c{C}engelk\"{o}y 81220, \.{I}stanbul, Turkey}

\texttt{cemsinan@gursey.gov.tr}

{\vskip1.0cm}

\textbf{Abstract}

{\vskip0.5cm}
\end{center}

All possible interactions of a point particle with background
electromagnetic, gravitational and higher-spin fields is considered in the
two-time physics worldline formalism in (d,2) dimensions. This system has a
counterpart in a recent formulation of two-time physics in non-commutative
field theory with local Sp(2) symmetry. In either the worldline or field
theory formulation, a general Sp(2) algebraic constraint governs the
interactions, and determines equations that the background fields of any
spin must obey. The constraints are solved in the classical worldline
formalism ($\hbar \rightarrow 0$ limit) as well as in the field theory
formalism (all powers of $\hbar $). The solution in both cases coincide for
a certain 2T to 1T holographic image which describes a relativistic particle
interacting with background fields of any spin in (d-1,1) dimensions. Two
disconnected branches of solutions exist, which seem to have a
correspondence as massless states in string theory, one containing low spins
in the zero Regge slope limit, and the other containing high spins in the
infinite Regge slope limit. \newpage \pagenumbering{arabic}

\section{Introduction}

Local Sp$\left( 2\right) $ symmetry, and its supersymmetric generalization,
is the principle behind two-time physics (2TPhysics) \cite{survey2T}-\cite
{ncsp}. For a spinless particle, in the worldline formalism with local Sp$%
\left( 2\right) $ symmetry, the action is
\begin{equation}
S=\int d\tau \left( \partial _{\tau }X^{M}P_{M}-A^{ij}Q_{ij}\left(
X,P\right) \right) ,  \label{worldlineaction}
\end{equation}
where $A^{ij}\left( \tau \right) =A^{ji}\left( \tau \right) $ with $i,j=1,2,$
is the Sp$\left( 2\right) $ gauge potential. This action is local Sp$\left(
2\right) $ invariant (see \cite{emgrav} and below) provided the three $%
Q_{ij}\left( X,P\right) $ are any general phase space functions that satisfy
the Sp$\left( 2\right) $ algebra under Poisson brackets
\begin{equation}
\left\{ Q_{ij},Q_{kl}\right\} =\varepsilon _{jk}Q_{il}+\varepsilon
_{ik}Q_{jl}+\varepsilon _{jl}Q_{ik}+\varepsilon _{il}Q_{jk}.  \label{poisson}
\end{equation}
The antisymmetric $\varepsilon _{ij}=-\varepsilon _{ji}$ is the invariant
metric of Sp$\left( 2\right) $ that is used to raise/lower indices.

The goal of this paper is to determine all possible $Q_{ij}$ as functions of
phase space $X^{M},P^{M}$ that satisfy this algebra. The solution will be
given in the form of a power series in momenta which identify the background
fields $A_{M}\left( X\right) $ and $Z_{ij}^{M_{1}M_{2}\cdots M_{s}}\left(
X\right) $
\begin{equation}
Q_{ij}\left( X,P\right) =\sum_{s=0}^{\infty }Z_{ij}^{M_{1}M_{2}\cdots
M_{s}}\left( X\right) \,\left( P_{M_{1}}+A_{M_{1}}\left( X\right) \right)
\left( P_{M_{2}}+A_{M_{2}}\left( X\right) \right) \cdots \left(
P_{M_{s}}+A_{M_{s}}\left( X\right) \right) .  \label{expand}
\end{equation}
These $d+2$ dimensional fields will describe the particle interactions with
the Maxwell field, gravitational field, and higher-spin fields, when
interpreted in $d$ dimensions as a particular 2T to 1T holographic $d$%
-dimensional picture of the higher $d+2$ dimensional theory. Furthermore, as
is the usual case in 2Tphysics, there are a large number of 2T to 1T
holographic $d$-dimensional pictures of the same $d+2$ dimensional system.
For any fixed background the resulting $d$-dimensional dynamical systems are
interpreted as a unified family of 1T dynamical systems that are related to
each other by duality type Sp$\left( 2\right) $ transformations. This latter
property is one of the novel unification features offered by 2Tphysics.

In previous investigations the general solution up to maximum spin $s=2$ was
determined \cite{emgrav}. The most general solution with higher-spin fields
for arbitrary spin $s$ is given here. Such $Q_{ij}\left( X,P\right) $ are
then the generators of local Sp$\left( 2\right) $, with transformations of
the coordinates $\left( X^{M},P_{M}\right) $ given by
\begin{equation}
\delta _{\omega }X^{M}=-\omega ^{ij}\left( \tau \right) \frac{\partial Q_{ij}%
}{\partial P_{M}},\quad \delta _{\omega }P^{M}=\omega ^{ij}\left( \tau
\right) \frac{\partial Q_{ij}}{\partial X_{M}},  \label{spgauge1}
\end{equation}
where the $\omega ^{ij}\left( \tau \right) =\omega ^{ji}\left( \tau \right) $
are the local Sp$\left( 2\right) $ gauge parameters \cite{emgrav}. When the
Sp$\left( 2\right) $ gauge field $A^{ij}\left( \tau \right) $ transforms as
usual in the adjoint representation,
\begin{equation}
\delta _{\omega }A^{ij}=\partial _{\tau }\omega ^{ij}+\left[ \omega ,A\right]
^{ij},  \label{spgauge2}
\end{equation}
the action (\ref{worldlineaction}) is gauge invariant $\delta _{\omega }S=0,$
provided the background fields $Z_{ij}^{M_{1}M_{2}\cdots M_{s}}\left(
X\right) $ and $A_{M}\left( X\right) $ are such that $Q_{ij}\left(
X,P\right) $ satisfy the Sp$\left( 2\right) $ algebra above. Thus, through
the requirement of local Sp$\left( 2\right) ,$ the background fields are
restricted by certain differential equations that will be derived and solved
in this paper. As we will see, the solution permits certain unrestricted
functions that are interpreted as background fields of any spin in \textit{%
two lower dimensions}.

An important aspect for the physical interpretation is that the equations of
motion for the gauge field $A^{ij}\left( \tau \right) $ restricts the system
to phase space configurations that obey the classical on-shell condition
\begin{equation}
Q_{ij}\left( X,P\right) =0.  \label{singlet}
\end{equation}
The meaning of this equation is that physical configurations are gauge
invariant and correspond to singlets of Sp($2).$ These equations have an
enormous amount of information and provide a unification of a large number
of one-time physics systems in the form of a single higher dimensional
theory. One-time dynamical systems appear then as holographic images of the
unifying bulk system that exists in one extra timelike and one extra
spacelike dimensions. We will refer to this property as 2T to 1T holography.

This holography comes about because there are non-trivial solutions to (\ref
{singlet}) only iff the spacetime includes two timelike dimensions with
signature $\left( d,2\right) $. By Sp$\left( 2\right) $ gauge fixing, two
dimensions are eliminated, and $d$ dimensions are embedded inside $d+2$
dimensions in ways that are distinguishable from the point of view of the
remaining timelike dimension. This provides the holographic images that are
interpreted as distinguishable one-time dynamics. Thus one obtains a
multitude of non-trivial solutions with different physical interpretations
from the point of view of one time physics. Hence, for each set of fixed
background fields that obey the local Sp$\left( 2\right) $ conditions, the
2Tphysics action above unifies various one-time physical systems (i.e. their
actions, equations of motion, etc.) into a single 2Tphysics system.

To find all possible actions, one must first find all possible solutions of $%
Q_{ij}\left( X,P\right) $ that satisfy the \textit{off-shell Sp}$\left(
2\right) $ \textit{algebra before imposing the singlet condition}.

The simplest example is given by \cite{conf}
\begin{equation}
Q_{11}=X\cdot X,\quad Q_{12}=X\cdot P,\quad Q_{22}=P\cdot P.
\label{simplest}
\end{equation}
This form satisfies the Sp$\left( 2\right) $ algebra for any number of
dimensions $X^{M},P^{M},$ $M=1,2,\cdots D,$ and any signature for the flat
metric $\eta _{MN}$ used in the dot products $X\cdot P=X^{M}P^{N}\eta _{MN},
$ etc. However, the on-shell condition (\ref{singlet}) has non-trivial
solutions only and only if the metric $\eta _{MN}$ has signature $\left(
d,2\right) $ with two time-like dimensions: if the signature were Euclidean
the solutions would be trivial $X^{M}=P^{M}=0;$ if there would be only one
timelike dimension, then there would be no angular momentum $L^{MN}=0$ since
$X^{M},P^{M}$ would both be lightlike and parallel to each other; and if
there were more than two timelike dimensions the solutions would have ghosts
that could not be removed by the available Sp$\left( 2\right) $ gauge
symmetry. Hence two timelike dimensions is the only non-trivial physical
case allowed by the Sp$\left( 2\right) $ singlet condition (i.e. gauge
invariance).

The general classical worldline problem that we will solve in this paper has
a counterpart in non-commutative field theory (NCFT) with local Sp$\left(
2\right) $ symmetry as formulated recently in \cite{ncsp}. The solution that
we give here provides also a solution to the non-commutative (NC) field
equations of motions that arise in that context. We note that in NCFT the
same field $Q_{ij}\left( X,P\right) $ emerges as the local Sp$\left(
2\right) $ covariant left-derivative including the gauge field. The field
strength is given by
\begin{equation}
G_{ij,kl}=\left[ Q_{ij},Q_{kl}\right] _{\star }-i\hbar \left( \varepsilon
_{jk}Q_{il}+\varepsilon _{ik}Q_{jl}+\varepsilon _{jl}Q_{ik}+\varepsilon
_{il}Q_{jk}\right) ,  \label{fijkl}
\end{equation}
where the star commutator $\left[ Q_{ij},Q_{kl}\right] _{\star }=Q_{ij}\star
Q_{kl}-Q_{kl}\star Q_{ij}$ is constructed using the Moyal star product
\begin{equation}
Q_{ij}\star Q_{kl}=\left. \exp \left( \frac{i\hbar }{2}\eta ^{MN}\left(
\frac{\partial }{\partial X^{M}}\frac{\partial }{\partial \tilde{P}^{N}}-%
\frac{\partial }{\partial P^{M}}\frac{\partial }{\partial \tilde{X}^{N}}%
\right) \right) Q_{ij}\left( X,P\right) Q_{kl}\left( \tilde{X},\tilde{P}%
\right) \right| _{X=\tilde{X},P=\tilde{P}}.  \label{starpr}
\end{equation}
Although there are general field configurations in NCFT that include
non-linear field interactions \cite{ncsp}, we will concentrate on a special
solution of the NCFT equations of motion. The special solution is obtained
when $G_{ij,kl}=0,$ (i.e. $Q_{ij}$ satisfies the Sp$\left( 2\right) $
algebra under star commutators), and $Q_{ij}$ annihilates a wavefunction $%
\Phi \left( X,P\right) $ that is interpreted as a singlet of Sp$\left(
2\right) $%
\begin{eqnarray}
\frac{1}{i\hbar }\left[ Q_{ij},Q_{kl}\right] _{\star } &=&\varepsilon
_{jk}Q_{il}+\varepsilon _{ik}Q_{jl}+\varepsilon _{jl}Q_{ik}+\varepsilon
_{il}Q_{jk}  \label{ncft1} \\
Q_{ij}\star \Phi &=&0.  \label{ncft2}
\end{eqnarray}
These field equations are equivalent to the first quantization of the
worldline theory in a quantum phase space formalism (as opposed to the more
traditional pure position space or pure momentum space formalism).

Compared to the Poisson bracket that appears in (\ref{poisson}) the star
commutator is an infinite series in powers of $\hbar $. It reduces to the
Poisson bracket in the classical limit $\hbar \rightarrow 0,$
\begin{equation}
\frac{1}{i\hbar }\left[ Q_{ij},Q_{kl}\right] _{\star }\rightarrow \left\{
Q_{ij},Q_{kl}\right\} .
\end{equation}
Therefore, any solution for $Q_{ij}\left( X,P\right) $ of the form (\ref
{expand}) that satisfies the Poisson bracket Sp$\left( 2\right) $ algebra (%
\ref{poisson}) is normally expected to be only an approximate semi-classical
solution of the NCFT equations (\ref{ncft1},\ref{ncft2}) that involve the
star product (\ref{starpr}). However, we find a much better than expected
solution: by choosing certain gauges of the Sp$\left( 2\right) $ gauge
symmetry in the NCFT approach, we learn that the classical solution of the
Poisson bracket algebra in eq.(\ref{poisson}) is also an exact solution of
the star commutator algebra (\ref{ncft1}) to all orders of $\hbar $.

We will see that the solution has two disconnected branches of background
fields. The first branch has only low spins $s\leq 2$ including the
gravitational field. The second branch has only high spins $s\geq 2$
starting with the gravitational field. These appear to have a correspondence
to the massless states in string theory at extreme limits of the string
tension $T\sim 1/\alpha ^{\prime }\rightarrow 0,\infty .$ Indeed when the
Regge slope $\alpha ^{\prime }$ goes to zero by fixing the graviton state
only the low spin $s\leq 2$ massless states survive, and when the Regge
slope $\alpha ^{\prime }$ goes to infinity there are an infinite number of
high spin massless states. The high spin fields that we find here correspond
to those massless states obtained from the graviton trajectory $s\geq 2$.

The paper is organized as follows. In section 2 we discuss an infinite
dimensional canonical transformation symmetry of the equations. In section
3, we use the symmetry to simplify the content of the background fields that
appear in the expansion (\ref{expand}), and we impose the Poisson bracket
algebra (\ref{poisson}) to determine the field equations that must be
satisfied by the remaining background fields. In section 4, we discuss two
special coordinate systems to solve the field equations. Then we impose the
gauge invariance condition $Q_{ij}=0,$ and interpret the holographic image
as a relativistic particle in $d$ dimensions $x^{\mu },$ moving in the
background of fields of various spins, including a scalar field $u\left(
x\right) ,$ gauge field $A_{\mu }\left( x\right) ,$ gravitational field $%
g^{\mu \nu }\left( x\right) $ and higher-spin fields $g^{\mu _{1}\mu
_{2}\cdots \mu _{s}}\left( x\right) $ for any spin $s$. We also derive the
gauge transformation rules of the higher-spin fields in $d$ dimensions, and
learn that there are two disconnected branches. In section 5 we show that
the classical solution is also an exact quantum solution of the star product
system that emerges in NCFT with local Sp$\left( 2\right) $ symmetry. In
section 6, we conclude with some remarks.

Our 2T approach to higher-spin fields makes connections to other methods in
the literature. One connection occurs in a special Sp$\left( 2\right) $
gauge (section 4) which links to Dirac's formulation of SO$\left( d,2\right)
$ conformal symmetry by using $d+2$ dimensional fields to represent $d$
dimensional fields \cite{dirac}. In our paper this method is extended to all
high spin fields as a particular 2T to 1T holographic picture. The $d$%
-dimensional system of this holographic picture has an overlap with a
description of higher-spin fields given in \cite{segal}, which is probably
related to the approach of Vasiliev \textit{et al.} (see \cite{vasiliev} and
references therein). It was shown in \cite{segal} that our special 1T
holographic picture, when translated to the second order formalism (as
opposed to the phase space formalism), is a completion of the de
Wit-Freedman action \cite{dwitf} for a spinless relativistic particle
interacting with higher-spin background fields.

\section{Infinite dimensional symmetry}

Although our initial problem is basically at the classical level, we will
adopt the idea of the associative star product, in the $\hbar \rightarrow 0$
limit, as a convenient formalism. In this way our discussion will be
naturally extended in section 5 to the case of NCFT which will be valid for
any $\hbar $. The Poisson bracket is written in terms of the star product (%
\ref{starpr}) as a limit of the form
\begin{equation}
\left\{ A,B\right\} =\lim_{\hbar \rightarrow 0}\frac{1}{i\hbar }\left(
A\star B-B\star A\right) =\frac{\partial A}{\partial X^{M}}\frac{\partial B}{%
\partial P_{M}}-\frac{\partial B}{\partial X^{M}}\frac{\partial A}{\partial
P_{M}}.
\end{equation}
Consider the following transformation of any function of phase space $%
A\left( X,P\right) $%
\begin{eqnarray}
A\left( X,P\right) &\rightarrow &\tilde{A}\left( X,P\right) =\lim_{\hbar
\rightarrow 0}e^{-i\varepsilon /\hbar }\star A\star e^{i\varepsilon /\hbar }
\\
&=&A+\left\{ \varepsilon ,A\right\} +\frac{1}{2!}\left\{ \varepsilon
,\left\{ \varepsilon ,A\right\} \right\} +\frac{1}{3!}\left\{ \varepsilon
,\left\{ \varepsilon ,\left\{ \varepsilon ,A\right\} \right\} \right\}
+\cdots
\end{eqnarray}
for any $\varepsilon \left( X,P\right) .$ If the $\hbar \rightarrow 0$ is
not applied, every Poisson bracket on the right hand side is replaced by the
star commutator. It is straightforward to see that the Poisson bracket of
two such functions transforms as
\begin{eqnarray}
\left\{ A,B\right\} &\rightarrow &\left\{ \tilde{A},\tilde{B}\right\}
=\left\{ \left( \lim_{\hbar \rightarrow 0}e^{-i\varepsilon /\hbar }\star
A\star e^{i\varepsilon /\hbar }\right) ,\left( \lim_{\hbar \rightarrow
0}e^{-i\varepsilon /\hbar }\star B\star e^{i\varepsilon /\hbar }\right)
\right\} \\
&=&\lim_{\hbar \rightarrow 0}e^{-i\varepsilon /\hbar }\star \left\{
A,B\right\} \star e^{i\varepsilon /\hbar }.  \label{transpoisson}
\end{eqnarray}
In particular, the phase space variables $X^{M},P_{M}$ transform into
\begin{equation}
\tilde{X}^{M}=\lim_{\hbar \rightarrow 0}e^{-i\varepsilon /\hbar }\star
X^{M}\star e^{i\varepsilon /\hbar },\quad \tilde{P}_{M}=\lim_{\hbar
\rightarrow 0}e^{-i\varepsilon /\hbar }\star P_{M}\star e^{i\varepsilon
/\hbar },  \label{canonic}
\end{equation}
and one can easily verify that the canonical Poisson brackets remain
invariant
\begin{equation}
\left\{ X^{M},P_{N}\right\} =\left\{ \tilde{X}^{M},\tilde{P}_{N}\right\}
=\delta _{N}^{M}.
\end{equation}
So, the transformation we have defined is the most general canonical
transformation. In particular, for infinitesimal $\varepsilon ,$ one has
\begin{equation}
\delta _{\varepsilon }X^{M}=-\partial \varepsilon /\partial P_{M},\quad
\delta _{\varepsilon }P_{M}=\partial \varepsilon /\partial X^{M},
\label{canon}
\end{equation}
which is again recognized as a general canonical transformation with
generator $\varepsilon \left( X,P\right) .$ The generator $\varepsilon
\left( X,P\right) $ contains an infinite number of parameters, so this set
of transformations form an infinite dimensional group. There is a
resemblance between (\ref{canon}) and the expressions in (\ref{spgauge1})
but note that those include general $\tau $ dependence in $\omega
^{ij}\left( \tau \right) $ and therefore are quite different.

Under general canonical transformations (\ref{canonic}) the particle action (%
\ref{worldlineaction}) transforms as
\begin{equation}
S_{Q_{ij}}\left( X,P\right) \rightarrow S_{Q_{ij}}\left( \tilde{X},\tilde{P}%
\right) =\int d\tau \left( \partial _{\tau }\tilde{X}^{M}\tilde{P}%
_{M}-A^{ij}Q_{ij}\left( \tilde{X},\tilde{P}\right) \right) .
\end{equation}
The first term is invariant $\int d\tau \left( \partial _{\tau }\tilde{X}^{M}%
\tilde{P}_{M}\right) =\int d\tau \left( \partial _{\tau }X^{M}P_{M}\right) .$
This is easily verified for infinitesimal $\varepsilon \left( X,P\right) $
since $\delta _{\varepsilon }\left( \partial _{\tau }X^{M}P_{M}\right) $ is
a total derivative
\[
\delta _{\varepsilon }\left( \partial _{\tau }X^{M}P_{M}\right) =\partial
_{\tau }\left( \varepsilon -P\cdot \frac{\partial \varepsilon }{\partial P}%
\right) .
\]
However, the full action (\ref{worldlineaction}) is not invariant. Instead,
it is mapped to a new action with a new set of background fields $\tilde{Q}%
_{ij}\left( X,P\right) $ given by
\begin{eqnarray}
Q_{ij}\left( \tilde{X},\tilde{P}\right) &=&\tilde{Q}_{ij}\left( X,P\right)
=\lim_{\hbar \rightarrow 0}e^{-i\varepsilon /\hbar }\star Q_{ij}\left(
X,P\right) \star e^{i\varepsilon /\hbar }  \label{Qprime} \\
S_{Q_{ij}}\left( X,P\right) &\rightarrow &S_{Q_{ij}}\left( \tilde{X},\tilde{P%
}\right) =S_{\tilde{Q}_{ij}}\left( X,P\right) =\int d\tau \left( \partial
_{\tau }X^{M}P_{M}-A^{ij}\tilde{Q}_{ij}\left( X,P\right) \right) .
\end{eqnarray}
After taking into account (\ref{transpoisson}) we learn that the new $\tilde{%
Q}_{ij}\left( X,P\right) $ also satisfies the Sp$\left( 2\right) $ algebra (%
\ref{poisson}) if the old one $Q_{ij}\left( X,P\right) $ did. Thus, the new
action $S_{\tilde{Q}_{ij}}\left( X,P\right) $ is again invariant under the
local Sp$\left( 2\right) $ symmetry (\ref{spgauge1},\ref{spgauge2}) using
the new generators $\tilde{Q}_{ij}$, and is therefore in the class of
actions we are seeking.

Thus, if we find a given solution for the background fields $%
Z_{ij}^{M_{1}M_{2}\cdots M_{s}}\left( X\right) $ and $A_{M}\left( X\right) $
in (\ref{expand}) such that $Q_{ij}\left( X,P\right) $ satisfy the Sp$\left(
2\right) $ algebra (\ref{poisson}), we can find an infinite number of new
solutions $\tilde{Z}_{ij}^{M_{1}M_{2}\cdots M_{s}}\left( X\right) $ and $%
\tilde{A}_{M}\left( X\right) $ by applying to $Q_{ij}\left( X,P\right) $ the
infinite dimensional canonical transformation (\ref{Qprime}) for any
function of phase space $\varepsilon \left( X,P\right) .$ We may write this
function in a series form similar to (\ref{expand}) to display its infinite
number of local parameters $\varepsilon _{s}^{M_{1}M_{2}\cdots M_{s}}(X)$
\begin{equation}
\varepsilon (X,P)=\sum_{s=0}^{\infty }\varepsilon _{s}^{M_{1}M_{2}\cdots
M_{s}}\left( X\right) \,(P_{M_{1}}+A_{M_{1}})(P_{M_{2}}+A_{M_{2}})\cdots
(P_{M_{s}}+A_{M_{s}}).  \label{epsilon}
\end{equation}
Although this set of transformations is not a symmetry of the worldline
action for a fixed set of background fields, it is evidently a symmetry in
the space of actions for all possible background fields, by allowing those
fields to transform. It is also an automorphism symmetry of the Sp$\left(
2\right) $ algebra (\ref{poisson}), and of the on-shell singlet condition (%
\ref{singlet}) which identifies the physical sector. Furthermore, in the
NCFT setting of \cite{ncsp} this is, in fact, the local Sp$\left( 2\right) $
symmetry with $\varepsilon (X,P)$ playing the role of the local gauge
parameter in NC gauge field theory. We will use this information to simplify
our task of finding the general solution.

\section{Imposing the Poisson bracket algebra}

By taking into account the infinite dimensional symmetry of the previous
section, we can always map a general $Q_{11}\left( X,P\right) $ to a
function of only $X^{M}$%
\begin{equation}
\tilde{Q}_{11}=\lim_{\hbar \rightarrow 0}e^{-i\varepsilon /\hbar }\star
Q_{ij}\left( X,P\right) \star e^{i\varepsilon /\hbar }=W\left( X\right) .
\label{W1}
\end{equation}
Conversely, given $\tilde{Q}_{11}=W\left( X\right) $ we may reconstruct the
general $Q_{ij}\left( X,P\right) $ by using the inverse transformation
\begin{equation}
Q_{11}\left( X,P\right) =\lim_{\hbar \rightarrow 0}e^{i\varepsilon /\hbar
}\star W\left( X\right) \star e^{-i\varepsilon /\hbar }.  \label{W2}
\end{equation}
There is enough symmetry to map $W\left( X\right) $ to any desired non-zero
function of $X^{M}$ that would permit the reconstruction (\ref{W2}) of the
general $Q_{11}\left( X,P\right) ,$ but we postpone this freedom until a
later stage (see next section) in order to first exhibit a more general
setting.

After fixing $\tilde{Q}_{11}=W\left( X\right) ,$ there is a remaining
subgroup of transformations $\varepsilon \left( X,P\right) $ for which the
Poisson bracket $\left\{ \varepsilon ,W\right\} $ vanishes, and hence $%
W\left( X\right) $ is invariant under it. Using the form of (\ref{epsilon})
we see that the subgroup corresponds to those transformations that satisfy
the condition
\begin{equation}
\frac{\partial W}{\partial X^{M_{1}}}\varepsilon _{s\geq
1}^{M_{1}M_{2}\cdots M_{s}}\left( X\right) =0,\quad any\,\,\varepsilon
_{0}\left( X\right) .  \label{subgroup}
\end{equation}
This subgroup can be used to further simplify the problem. To see how, let
us consider the expansions
\begin{eqnarray}
\tilde{Q}_{11}\left( X,P\right) &=&W\left( X\right) \\
\tilde{Q}_{12}\left( X,P\right) &=&\sum_{s=0}^{\infty
}V_{s}^{M_{1}M_{2}\cdots M_{s}}\,\left( P_{M_{1}}+A_{M_{1}}\right) \left(
P_{M_{2}}+A_{M_{2}}\right) \cdots \left( P_{M_{s}}+A_{M_{s}}\right) \\
\tilde{Q}_{22}\left( X,P\right) &=&\sum_{s=0}^{\infty
}G_{s}^{M_{1}M_{2}\cdots M_{s}}\,\,\left( P_{M_{1}}+A_{M_{1}}\right) \left(
P_{M_{2}}+A_{M_{2}}\right) \cdots \left( P_{M_{s}}+A_{M_{s}}\right)
\end{eqnarray}
where $G_{s}^{M_{1}M_{2}\cdots M_{s}}\left( X\right) $ and $%
V_{s}^{M_{1}M_{2}\cdots M_{s}}\left( X\right) $ are fully symmetric local
tensors of rank $s.$

We can completely determine the coefficients $V_{s}^{M_{1}M_{2}\cdots M_{s}}$
in terms of $W$ and $G_{s}^{M_{1}M_{2}\cdots M_{s}}\,$\ by imposing one of
the Sp$\left( 2\right) $ conditions $\left\{ \tilde{Q}_{11},\tilde{Q}%
_{22}\right\} =4\tilde{Q}_{12}$
\begin{equation}
V_{s-1}^{M_{2}M_{3}\cdots M_{s}}=\frac{s}{4}\frac{\partial W}{\partial
X^{M_{1}}}G_{s}^{M_{1}M_{2}\cdots M_{s}}.  \label{vsm1}
\end{equation}
Furthermore, by imposing another Sp$\left( 2\right) $ condition $\left\{
\tilde{Q}_{11},\tilde{Q}_{12}\right\} =2\tilde{Q}_{11}$ we find
\begin{equation}
\frac{\partial W}{\partial X^{M_{1}}}V_{1}^{M_{1}}=2W,\quad \frac{\partial W%
}{\partial X^{M_{1}}}V_{s\geq 2}^{M_{1}M_{2}\cdots M_{s}}=0.
\label{Vconditions}
\end{equation}
Now, for such $V_{s}^{M_{1}M_{2}\cdots M_{s}},$ by using the remaining
subgroup symmetry (\ref{subgroup}) we can transform to a frame in which all $%
V_{s}^{M_{1}M_{2}\cdots M_{s}}$ for $s\geq 2$ vanish. By comparing the
expressions (\ref{subgroup},\ref{Vconditions}) and counting parameters we
see that this must be possible. To see it in more detail, we derive the
infinitesimal transformation law for $G_{s}^{M_{1}M_{2}\cdots M_{s}}$ and $%
A_{M}$ from
\begin{equation}
\delta Q_{22}=\left( \partial Q_{22}/\partial A\right) \delta
A+\sum_{s}\left( \partial Q_{22}/\partial G_{s}\right) \delta G_{s}=\left\{
\varepsilon ,Q_{22}\right\}   \label{deltaQ22}
\end{equation}
by expanding both sides in powers of $\left( P+A\right) $ and comparing
coefficients. We write the result in symbolic notation by suppressing the
indices
\begin{eqnarray}
\delta A_{M} &=&\partial _{M}\tilde{\varepsilon}_{0}-\mathcal{L}%
_{\varepsilon _{1}}A_{M},\quad   \label{deltaG1} \\
\delta G_{0} &=&-\varepsilon _{1}\cdot \partial G_{0}, \\
\delta G_{1} &=&-\mathcal{L}_{\varepsilon _{1}}G_{1}-\left( \varepsilon
_{2}\cdot \partial G_{0}+\varepsilon _{2}FG_{1}\right) , \\
\delta G_{s\geq 2} &=&-\mathcal{L}_{\varepsilon _{1}}G_{s}-\left(
\varepsilon _{2}\cdot \partial G_{s-1}-G_{s-1}\cdot \partial \varepsilon
_{2}+\varepsilon _{2}FG_{s}\right) -\cdots   \nonumber \\
&&\cdots -\left( \varepsilon _{s}\cdot \partial G_{1}-G_{1}\cdot \partial
\varepsilon _{s}+\varepsilon _{s}FG_{2}\right) -\left( \varepsilon
_{s+1}\cdot \partial G_{0}+\varepsilon _{s+1}FG_{1}\right) .  \label{deltaGs}
\end{eqnarray}
In $\delta G_{s\geq 2}$ the dots $\cdots $ represent terms of the form $%
\left( \varepsilon _{k}\cdot \partial G_{s-k+1}-G_{s-k+1}\cdot \partial
\varepsilon _{k}+\varepsilon _{k}FG_{s-k+2}\right) $ for all $2<k<s.$ Here $%
F_{MN}$ is the gauge field strength
\begin{equation}
F_{MN}\left( X\right) =\partial _{M}A_{N}\left( X\right) -\partial
_{N}A_{M}\left( X\right) ,
\end{equation}
$\mathcal{L}_{\varepsilon _{1}}G_{s}$ is the Lie derivative of the tensor $%
G_{s}^{M_{1}M_{2}\cdots M_{s}}$ with respect to the vector $\varepsilon
_{1}^{M}$%
\begin{equation}
\left( \mathcal{L}_{\varepsilon _{1}}G_{s}\right) ^{M_{1}M_{2}\cdots
M_{s}}=\varepsilon _{1}\cdot \partial G_{s}^{M_{1}M_{2}\cdots
M_{s}}-\partial _{K}\varepsilon _{1}^{M_{1}}G_{s}^{KM_{2}\cdots
M_{s}}-\cdots -\partial _{K}\varepsilon _{1}^{M_{s}}G_{s}^{M_{1}M_{2}\cdots
K}.  \label{lie}
\end{equation}
In the other terms, $\varepsilon _{k}\cdot \partial G_{l}$ (similarly $%
G_{k}\cdot \partial \varepsilon _{l}$) is the tensor
\begin{equation}
\varepsilon _{k}\cdot \partial G_{l}=\frac{k!l!}{\left( k+l-1\right) !}%
\varepsilon _{k}^{M_{1}(M_{2}\cdots M_{k}}\partial
_{M_{1}}G_{l}^{M_{k+1}\cdots M_{k+l})},
\end{equation}
where all un-summed upper indices ($k+l-1$ of them) are symmetrized, and $%
\varepsilon _{k}FG_{l}$ is the tensor
\begin{equation}
\varepsilon _{k}FG_{l}=\frac{k!l!}{\left( k+l-2\right) !}\varepsilon
_{k}^{M_{1}(M_{2}\cdots M_{k}}G_{l}^{M_{k+1}\cdots )M_{k+l}}F_{M_{1}M_{k+l}},
\end{equation}
where all un-summed upper indices ($k+l-2$ of them) are symmetrized. Finally
$\tilde{\varepsilon}_{0}$ which appears in $\delta A$ is defined by $\tilde{%
\varepsilon}_{0}=\varepsilon _{0}+\varepsilon _{1}\cdot A$ .

From $\delta A_{M}$ it is evident that $\tilde{\varepsilon}_{0}\left(
X\right) $ is a Yang-Mills type gauge parameter, and from $\mathcal{L}%
_{\varepsilon _{1}}G_{s}$ it is clear that $\varepsilon _{1}^{M}\left(
X\right) $ is the parameter of general coordinate transformations in
position space. The remaining parameters $\varepsilon _{s\geq 2}\left(
X\right) $ are gauge parameters for high spin fields (note that the
derivative of the $\varepsilon _{s}$ appear in the transformation rules).
From the transformation laws for $\delta A,\delta G_{s}$ we find the
transformation law for $\delta V_{s}^{M_{1}M_{2}\cdots M_{s}}$ by
contracting both sides of the equation above with $\partial _{M}W\left(
X\right) $. After using the subgroup condition (\ref{subgroup}) and the
definition (\ref{vsm1}) we find
\begin{eqnarray}
\delta V_{0} &=&-\mathcal{L}_{\varepsilon _{1}}V_{0},\quad \\
\delta V_{1} &=&-\mathcal{L}_{\varepsilon _{1}}V_{1}-\left( \varepsilon
_{2}\cdot \partial V_{0}+\varepsilon _{2}FV_{1}\right) \\
\delta V_{s\geq 2} &=&-\mathcal{L}_{\varepsilon _{1}}V_{s}-\left(
\varepsilon _{2}\cdot \partial V_{s-1}-V_{s-1}\cdot \partial \varepsilon
_{2}+\varepsilon _{2}FV_{s}\right) -\cdots  \nonumber \\
&&\cdots -\left( \varepsilon _{s}\cdot \partial V_{1}-V_{1}\cdot \partial
\varepsilon _{s}+\varepsilon _{s}FV_{2}\right) -\left( \varepsilon
_{s+1}\cdot \partial V_{0}+\varepsilon _{s+1}FV_{1}\right) .
\end{eqnarray}
The form of $\delta V_{k}$ is similar to the form of $\delta G_{k}$ as might
be expected, since it can also be obtained from $\delta Q_{12}=\left\{
\varepsilon ,Q_{12}\right\} ,$ but we have derived it by taking into account
the restriction (\ref{vsm1}) and the subgroup condition (\ref{subgroup}).

For $V_{s}$ of the form (\ref{Vconditions}) the subgroup parameters are
sufficient to transform to a frame where $V_{s\geq 2}=0.$ Therefore, we may
always start from a frame of the form
\begin{eqnarray}
\tilde{Q}_{11}\left( X,P\right) &=&W\left( X\right) ;\quad V_{1}\cdot
\partial W=2W,  \label{tq11} \\
\tilde{Q}_{12}\left( X,P\right) &=&V_{0}+V_{1}^{M}\left( P_{M}+A_{M}\right)
;\quad V_{0}=\frac{1}{4}\partial _{N}W\,G_{1}^{N},\quad V_{1}^{M}=\frac{1}{2}%
\partial _{N}W\,G_{2}^{MN},  \label{tq12} \\
\tilde{Q}_{22}\left( X,P\right) &=&\sum_{s=0}^{\infty
}G_{s}^{M_{1}M_{2}\cdots M_{s}}\,\left( P_{M_{1}}+A_{M_{1}}\right) \left(
P_{M_{2}}+A_{M_{2}}\right) \cdots \left( P_{M_{s}}+A_{M_{s}}\right)
\label{tq22}
\end{eqnarray}
and transform to the most general solution via
\begin{equation}
Q_{ij}\left( X,P\right) =\lim_{\hbar \rightarrow 0}e^{i\varepsilon /\hbar
}\star \tilde{Q}_{ij}\left( X,P\right) \star e^{-i\varepsilon /\hbar }.
\label{complete}
\end{equation}
In $\tilde{Q}_{22}$ the term $G_{1}^{M}$ may be set equal to zero by
shifting $A_{M}\rightarrow A_{M}-\frac{1}{2}\left( G_{2}\right)
_{MN}G_{1}^{N}+\cdots ,$ and then redefining all other background fields.
Here we have assumed that the tensor $G_{2}^{MN}$ has an inverse $\left(
G_{2}\right) _{MN}$; in fact, as we will see soon, it will have the meaning
of a metric. Therefore, we will assume $G_{1}^{M}=0$ without any loss of
generality. In that case we see from (\ref{tq12}) that we must also have $%
V_{0}=0.$

It suffices to impose the remaining relations of the Sp$\left( 2\right) $
algebra in this frame. By comparing the coefficients of every power of $%
(P+A) $ in the condition $\left\{ \tilde{Q}_{12},\tilde{Q}_{22}\right\} =2%
\tilde{Q}_{22}$ we derive the following equations
\begin{equation}
V_{1}^{M}F_{MN}=0,\quad \mathcal{L}_{V_{1}}G_{s}=-2G_{s},  \label{lv1g}
\end{equation}
where $\mathcal{L}_{V_{1}}G_{s}$ is the Lie derivative with respect to the
vector $V_{1}$ (see (\ref{lie})). These, together with
\begin{equation}
V_{1}^{M}=\frac{1}{2}\partial _{N}W\,G_{2}^{MN},\quad V_{1}\cdot \partial
W=2W,\quad V_{0}=0,\quad G_{1}=0,\quad \partial W\cdot G_{s\geq 3}=0,
\label{v1w}
\end{equation}
that we used before, provide the complete set of equations that must be
satisfied to have a closure of the Sp$\left( 2\right) $ algebra. These
background fields, together with the background fields provided by the
general $\varepsilon \left( X,P\right) $ through eq.(\ref{complete}),
generalize the results of \cite{emgrav}, where only $A_{M},G_{0}$ and $%
G_{2}^{MN}$ had been included.

There still is remaining canonical symmetry that keeps the form of the above
$\tilde{Q}_{ij}$ unchanged. This is given by the subgroup of symmetries
associated with $\tilde{\varepsilon}_{0}\left( X\right) ,\,\varepsilon
_{1}^{M}\left( X\right) $ which have the meaning of local parameters for
Yang-Mills and general coordinate transformations, and also the higher-spin
symmetries that satisfy
\begin{equation}
\partial W\cdot \varepsilon _{s\geq 1}=0,\quad \mathcal{L}%
_{V_{1}}\varepsilon _{s\geq 1}=0,\quad \partial G_{0}\cdot \varepsilon
_{2}=0.  \label{remain}
\end{equation}
The conditions in (\ref{remain}) are obtained after setting $G_{1}=V_{s\geq
2}=0$ and $\delta G_{1}=\delta V_{s\geq 2}=0,$ as well as using (\ref
{subgroup}).

It is possible to go further in using the remaining $\varepsilon _{s}\left(
X\right) $ transformations, but this will not be necessary since the
physical content of the worldline system will be more transparent by using
the background fields $G_{s}$ and $A_{M}$ identified up to this stage.
However, we will return to the remaining symmetry at a later stage to
clarify its action on the fields, and thus discover that there are two
disconnected branches.

\section{Choosing coordinates and W(X)}

As mentioned in the beginning of the previous section the original $%
\varepsilon \left( X,P\right) $ transformations permits a choice for the
function $W\left( X\right) ,$ while the surviving $\varepsilon
_{1}^{M}\left( X\right) $ which is equivalent to general coordinate
transformations further permits a choice for the vector $V_{1}^{M}\left(
X\right) ,$ as long as it is consistent with the differential conditions
given above. Given this freedom we will explore two choices for $W\left(
X\right) $ and $V_{1}^{M}\left( X\right) $ in this section.

\subsection{SO(d,2) covariant $W\left( X\right) =X^{2}$}

We choose $W\left( X\right) $ and $V_{1}^{M}\left( X\right) $ as follows
\begin{equation}
W\left( X\right) =X^{2}=X^{M}X^{N}\eta _{MN},\quad V_{1}^{M}\left( X\right)
=X^{M},  \label{wx2}
\end{equation}
where $\eta _{MN}$ is the metric for SO$\left( d,2\right) .$ These coincide
with part of the simplest Sp$\left( 2\right) $ system (\ref{simplest}). We
cannot choose any other signature $\eta _{MN}$ since we already know that
the constraints $Q_{ij}\left( X,P\right) =0$ have solutions only when the
signature includes two timelike dimensions.

Using (\ref{lv1g},\ref{v1w}), the metric $G_{2}^{MN}\left( X\right) $ takes
the form
\begin{equation}
G_{2}^{MN}=\eta ^{MN}+h_{2}^{MN}\left( X\right) ,\quad X\cdot \partial
h_{2}^{MN}=0,\quad h_{2}^{MN}X_{N}=0.
\end{equation}
$G_{2}^{MN}$ is an invertible metric. The fluctuation $h_{2}^{MN}\left(
X\right) $ is any homogeneous function of degree zero and it is orthogonal
to $X_{N}.$

Using the $\varepsilon _{0}\left( X\right) $ gauge degree of freedom we work
in the axial gauge $X\cdot A=0,$ then the condition $X^{M}F_{MN}=0$ reduces
to
\begin{equation}
\left( X\cdot \partial +1\right) A_{M}=0,\quad X\cdot A=0.
\end{equation}
Therefore $A_{M}\left( X\right) $ is any homogeneous vector of degree $%
\left( -1\right) $ and it is orthogonal to $X_{M}.$ There still is remaining
gauge symmetry $\delta A_{M}=\partial _{M}\varepsilon _{0}$ provided $%
\varepsilon _{0}\left( X\right) $ is a homogeneous function of degree zero
\begin{equation}
X\cdot \partial \varepsilon _{0}=0.
\end{equation}

Similarly, the higher-spin fields in (\ref{lv1g},\ref{v1w}) satisfy
\begin{equation}
\left( X\cdot \partial -s+2\right) G_{s\geq 3}^{M_{1}M_{2}\cdots
M_{s}}=0,\quad X_{M_{1}}G_{s\geq 3}^{M_{1}M_{2}\cdots M_{s}}=0,
\end{equation}
These equations are easily solved by homogeneous tensors of degree $s-2$
that are orthogonal to $X_{N}.$

The $\tilde{Q}_{ij}$ now take the SO$\left( d,2\right) $ covariant form
\begin{eqnarray}
\tilde{Q}_{11} &=&X^{2},\quad \tilde{Q}_{12}=X\cdot P, \\
\tilde{Q}_{22} &=&G_{0}+\sum_{s=2}^{\infty }G_{s}^{M_{1}\cdots M_{s}}\left(
P+A\right) _{M_{1}}\cdots \left( P+A\right) _{M_{s}}.
\end{eqnarray}
Thus, $Q_{11}$ and $Q_{12}$ are reduced to the form of the simplest
2Tphysics system (\ref{simplest}), while $Q_{22}$ contains the non-trivial
background fields. The remaining symmetry of (\ref{remain}) is given by
\begin{equation}
\partial G_{0}\cdot \varepsilon _{2}=0;\quad \left( X\cdot \partial
-s\right) \varepsilon _{s\geq 0}^{M_{1}M_{2}\cdots M_{s}}=0,\quad
X_{M_{1}}\varepsilon _{s\geq 1}^{M_{1}M_{2}\cdots M_{s}}=0.  \label{remaine}
\end{equation}
where all dot products involve the metric $\eta _{MN}$ of SO$\left(
d,2\right) .$ Hence the frame is SO$\left( d,2\right) $ covariant, and this
will be reflected in any of the gauge fixed versions of the theory. As
before, $\varepsilon _{0}\left( X\right) $ is the (homogeneous) Yang-Mills
type gauge parameter and the $\varepsilon _{s\geq 1}$ play the role of gauge
parameters for higher-spin fields as in (\ref{deltaGs}).

To solve the constraints $Q_{ij}=0$ we can choose various Sp$\left( 2\right)
$ gauges that produce the 2T to 1T holographic reduction. This identifies
some combination of the $X^{M}\left( \tau \right) $ with the $\tau $
parameter, thus reducing the 2Tphysics description to the 1Tphysics
description. Depending on the choice made, the 1T dynamics of the resulting
holographic picture in $d$ dimensions appears different from the point of
view of one-time. This produces various holographic pictures in an analogous
way to the free case discussed previously in \cite{survey2T}. We plan to
discuss several examples of holographic pictures in the presence of
background fields in a future publication.

\subsection{Lightcone type W(X)=-2$\protect\kappa w$}

There are coordinate choices that provide a shortcut to some of the
holographic pictures, although they do not illustrate the magical
unification of various 1T dynamics into a single 2T dynamics as clearly as
the SO$\left( d,2\right) $ formalism of the previous section. Nevertheless,
since such coordinate systems can be useful, we analyze one that is closely
related to the relativistic particle dynamics in $d$ dimensions\footnote{%
We call the coordinate system in this section ``lightcone type'' because, in
the Sp$\left( 2\right) $ gauge $\kappa =1,$ it can be related to a lightcone
type Sp$\left( 2\right) $ gauge ($X^{+^{\prime }}=1$) in the SO$\left(
d,2\right) $ covariant formalism of the previous section. Once the gauge is
fixed from either point of view, the 1T holographic picture describes the
massless relativistic particle (see e.g. \cite{survey2T}) including its
interactions with background fields.}. Following \cite{emgrav} we consider a
coordinate system $X^{M}=\left( \kappa ,w,x^{\mu }\right) $ and use the
symmetries to choose $V_{1}^{M}=(\kappa ,w,0)$ and $W=-2w\kappa $. Then, as
in \cite{emgrav} the solution for the gauge field, the spin-2 gravity field $%
G_{2}^{MN},$ and the scalar field $G_{0}$ are
\begin{eqnarray}
A_{\kappa } &=&-\frac{w}{2\kappa ^{2}\,}B\left( \frac{w}{\kappa },x\right)
,\quad A_{w}=\frac{1}{2\kappa }B\left( \frac{w}{\kappa },x\right) ,\quad
A_{\mu }=A_{\mu }\left( \frac{w}{\kappa },x\right) , \\
G_{2}^{MN} &=&\left(
\begin{array}{ccc}
\frac{\kappa }{w}\left( \gamma -1\right) & -\gamma & \frac{1}{\kappa }W^{\nu
} \\
-\gamma & \frac{w}{\kappa }\left( \gamma -1\right) & -\frac{w}{\kappa ^{2}}%
W^{\nu } \\
\frac{1}{\kappa }W^{\mu } & -\frac{w}{\kappa ^{2}}W^{\mu } & \frac{g^{\mu
\nu }}{\kappa ^{2}}
\end{array}
\right) ,  \label{G2} \\
G_{0} &=&\frac{1}{\kappa ^{2}}u\left( x,\frac{w}{\kappa }\right) .
\end{eqnarray}
where the functions $A_{\mu }\left( \frac{w}{\kappa },x\right) ,\,B\left(
\frac{w}{\kappa },x\right) ,\,\gamma \left( x,\frac{w}{\kappa }\right) ,$ $%
W^{\mu }\left( x,\frac{w}{\kappa }\right) ,$ $g^{\mu \nu }\left( x,\frac{w}{%
\kappa }\right) ,$ $u\left( x,\frac{w}{\kappa }\right) $ are arbitrary
functions of only $x^{\mu }$ and the ratio $\frac{w}{\kappa }$.

We now extend this analysis to the higher-spin fields. The equation
\begin{equation}
G_{s\geq 3}^{M_{1}M_{2}\cdots M_{s}}\cdot \partial _{M_{s}}W=0
\end{equation}
becomes
\begin{equation}
wG_{s\geq 3}^{M_{1}\cdots M_{s-1}\kappa }=-\kappa G_{s\geq 3}^{M_{1}\cdots
M_{s-1}w}.  \label{wgs}
\end{equation}
This shows that not all the components of $G_{s\geq 3}^{M_{1}M_{2}\cdots
M_{s}}$ are independent. The condition
\begin{equation}
\mathcal{L}_{V_{1}}G_{s}=-2G_{s}
\end{equation}
becomes
\begin{equation}
\left( \kappa \partial _{\kappa }+w\partial _{w}\right) G_{s}^{M_{1}\cdots
M_{s}}-\sum_{n=1}^{s}\delta _{\kappa }^{M_{n}}G_{s}^{M_{1}\cdots
M_{n-1}\kappa M_{n+1}\cdots M_{s}}-\sum_{n=1}^{s}\delta
_{w}^{M_{n}}G_{s}^{M_{1}\cdots M_{n-1}wM_{n+1}\cdots
M_{s}}=-2G_{s}^{M_{1}\cdots M_{s}}.
\end{equation}
Specializing the indices for independent components and also using the
relation (\ref{wgs}) between the components of $G_{s\geq
3}^{M_{1}M_{2}\cdots M_{s}}$ we get the solution for all components of $%
G_{s\geq 3}^{M_{1}M_{2}\cdots M_{s}}$ as
\begin{equation}
G_{s}^{\overbrace{\kappa \cdots \kappa }^{m}\overbrace{w\cdots w}^{n}\mu
_{1}\cdots \mu _{s-n-m}}=\kappa ^{m-2}\left( -w\right)
^{n}g_{s,(s-n-m)}^{\mu _{1}\cdots \mu _{s-n-m}},
\end{equation}
where $g_{s,k}^{\mu _{1}\mu _{2}\cdots \mu _{k}}\left( x,\frac{w}{\kappa }%
\right) ,$ with $k=1,\cdots ,s,$ are arbitrary functions and independent of
each other.

For this solution, the generators of Sp$\left( 2,R\right) $ in (\ref{tq11}-%
\ref{tq22}) become
\begin{eqnarray}
\tilde{Q}_{11} &=&-2\kappa w, \\
\tilde{Q}_{12} &=&\kappa p_{\kappa }+wp_{w}, \\
\tilde{Q}_{22} &=&-\frac{1}{\kappa w}\left[ \left( \kappa p_{\kappa }-\frac{%
wB}{2\kappa }\right) ^{2}+\left( wp_{w}+\frac{wB}{2\kappa }\right) ^{2}%
\right] +\frac{H+H^{\prime }}{\kappa ^{2}}.
\end{eqnarray}
where $H,H^{\prime },$ which contain the background fields, are defined by
\begin{eqnarray}
H &=&u+g^{\mu \nu }(p_{\mu }+A_{\mu })(p_{\nu }+A_{\nu })+\sum_{s=3}^{\infty
}g_{s,s}^{\mu _{1}\cdots \mu _{s}}(p_{\mu _{1}}+A_{\mu _{1}}) \cdots (p_{\mu
_{s}}+A_{\mu _{s}}),  \label{H} \\
H^{\prime } &=&\sum_{s=2}^{\infty }\sum_{k=0}^{s-1} g_{s,k}^{\mu _{1}\cdots
\mu _{k}}\,\left( \kappa p_{\kappa }-wp_{w}-\frac{wB}{\kappa }\right)
^{s-k}\,(p_{\mu _{1}}+A_{\mu _{1}})\cdots (p_{\mu _{k}}+A_{\mu _{k}}).
\end{eqnarray}
$H$ contains only the highest spin components $g_{s,s}^{\mu _{1}\mu
_{2}\cdots \mu _{s}}$ that emerge from $G_{s\geq 2}^{M_{1}M_{2}\cdots M_{s}}$%
. Here we have defined the metric $g^{\mu \nu }=g_{2,2}^{\mu \nu }$ as in (%
\ref{G2}). All the remaining lower spin components $g_{s,k}^{\mu _{1}\mu
_{2}\cdots \mu _{k}}$ with $k\leq s-1$ are included in $H^{\prime }.$ In the
$s=2$ term of $H^{\prime }$ we have defined $g_{2,0}\equiv \gamma \kappa /w$
and $g_{2,1}^{\mu }\equiv W^{\mu }$ in comparison to (\ref{G2})$.$ It can be
easily verified that these $\tilde{Q}_{ij}$ obey the Sp$(2,R)$ algebra for
any background fields $u,g_{\mu \nu },A_{\mu },B$ and $g_{s,k}^{\mu _{1}\mu
_{2}\cdots \mu _{k}}$ ($k=0,\cdots ,s$) that are \textit{arbitrary functions
of} $\left( x^{\mu },\frac{w}{\kappa }\right) $.

We next can choose some Sp$(2,R)$ \ gauges to solve the Sp$\left( 2,R\right)
$ constraints $\tilde{Q}_{ij}=0$ and reduce to a one-time theory containing
the higher-spin fields. As in the low spin-1 and spin-2 cases of \cite
{emgrav}, we choose $\kappa \left( \tau \right) =1$ and $p_{w}\left( \tau
\right) =0$, and solve $\tilde{Q}_{11}=\tilde{Q}_{12}=0$ in the form $%
w\left( \tau \right) =p_{\kappa }\left( \tau \right) =0$. We also use the
canonical freedom $\varepsilon _{0}$ to work in a gauge that insures $\frac{%
wB}{\kappa }\rightarrow 0,$ as $\frac{w}{\kappa }\rightarrow 0$. Then the $%
\tilde{Q}_{ij}$ simplify to
\begin{equation}
\tilde{Q}_{11}=\tilde{Q}_{12}=0,\quad \tilde{Q}_{22}=H,
\end{equation}
At this point, the two-time $d+2$ dimensional theory described by the
original action (\ref{worldlineaction}) reduces to a one-time theory in $d$
dimensions
\begin{equation}
S=\int d\tau \left( \partial _{\tau }x^{\mu }p_{\mu }-\frac{1}{2}%
A^{22}H\right) .  \label{actiond}
\end{equation}
This is a particular 2T to 1T holographic picture of the higher dimensional
theory obtained in a specific gauge. There remains unfixed one gauge
subgroup of Sp$\left( 2,R\right) $ which corresponds to $\tau $
reparametrization, and the corresponding Hamiltonian constraint is $H\sim 0$%
. There is also remaining canonical freedom which we will discuss below.
Here, in addition to the usual background fields $g_{\mu \nu }\left(
x\right) $, $A_{\mu }\left( x\right) $, $u\left( x\right) ,$ the Hamiltonian
includes the higher-spin fields $g_{s,s}^{\mu _{1}\mu _{2}\cdots \mu _{s}}$
that now are functions of only the $d$ dimensional coordinates $x^{\mu },$
since $w/\kappa =0$. Like $\gamma $ and $W^{\mu }$ in the gravity case, the
non-leading $g_{s,k}^{\mu _{1}\mu _{2}\cdots \mu _{k}}$ for $k<s$ decuple
from the dynamics that govern the time development of $x^{\mu }(\tau )$ in
this Sp$\left( 2\right) $ gauge.

A similar conclusion is obtained if we use the SO$\left( d,2\right) $
covariant formalism of the previous section when we choose the Sp$\left(
2\right) $ gauges $X^{+^{\prime }}=1,$ and $P^{+^{\prime }}=-P_{-^{\prime
}}=0$. The algebra for arriving at the final conclusion (\ref{actiond}) is
simpler in the coordinate frame of the present section\footnote{%
The $\left( \kappa ,w,x^{\mu }\right) $ coordinate system can be related to
the one in the previous section by a change of variables as follows.
Starting from the previous section define a lightcone type basis $X^{\pm
^{\prime }}=\left( X^{0^{\prime }}\pm X^{1^{\prime }}\right) /\sqrt{2}$, and
then make the change of variables $X^{+^{\prime }}=\kappa ,$ $X^{\mu
}=\kappa x^{\mu },$ $X^{-^{\prime }}=w+\kappa x^{2}/2$. Then $W=X\cdot
X=-2X^{+^{\prime }}X^{-^{\prime }}+X^{\mu }X_{\mu }=-2\kappa w$. The momenta
(with lower indices) are transformed as follows $P_{+^{\prime }}=p_{\kappa
}+p_{w}x^{2}/2-x\cdot p/\kappa ,$ $P_{-^{\prime }}=p_{w}$, and $P_{\mu
}=p_{\mu }/\kappa -p_{w}x_{\mu }.$ One can varify that $\dot{X}\cdot P=\dot{X%
}^{+^{\prime }}P_{+^{\prime }}+\dot{X}^{-^{\prime }}P_{-^{\prime }}+\dot{X}%
^{\mu }P_{\mu }=\dot{\kappa}p_{\kappa }+\dot{w}p_{w}+\dot{x}\cdot p$. In
this coordinate basis $X\cdot P=\kappa p_{\kappa }+wp_{w}$ and the dimension
operator $X\cdot \partial $ takes the form $X\cdot \partial =\kappa \partial
_{\kappa }+w\partial _{w}.$ This shows that all the results obtained with
the lightcone type $W=-2\kappa w$ can also be recovered from the covariant $%
W\left( X\right) =X^{2},$ and vice-versa.}, and this was the reason for
introducing the ``lightcone type'' $W=-2\kappa w$. However, from the SO$%
\left( d,2\right) $ covariant formalism we learn that there is a hidden SO$%
\left( d,2\right) $ in the $d$ dimensional action (\ref{actiond}). This can
be explored by examining the SO$\left( d,2\right) $ transformations produced
by $\varepsilon _{1}^{M}=\omega ^{MN}X_{N}$, obeying (\ref{remaine}), on all
the fields through the Lie derivative $\delta A_{M}=\mathcal{L}_{\varepsilon
_{1}}A_{M}$, $\delta G_{s}=-\mathcal{L}_{\varepsilon _{1}}G_{s},$ but this
will not be further pursued here.

In the present Sp$\left( 2\right) $ gauge we find a link to \cite{segal}
where the action (\ref{actiond}) was discussed. The symmetries inherited
from our $d+2$ dimensional approach (discussed below) have some overlap with
those discussed in \cite{segal}. It was shown in \cite{segal} that the first
order action (\ref{actiond}) improves and completes the second order action
discussed in \cite{dwitf}. Also, the incomplete local invariance discussed
in \cite{dwitf} is now completed by the inclusion of the higher powers of
velocity which were unknown in \cite{dwitf}. In the second order formalism
one verifies once more that the action describes a particle moving in the
background of arbitrary electromagnetic, gravitational and higher-spin
fields in the remaining $d$ dimensional spacetime.

\subsection{Surviving canonical symmetry in d dimensions}

Let us now analyze the form of the $d$ dimensional canonical symmetry
inherited from our $d+2$ dimensional approach. Recall that the infinite
dimensional canonical symmetry $\varepsilon \left( X,P\right) $ is not a
symmetry of the action, it is only a symmetry if the fields are permitted to
transform in the space of all possible worldline actions. What we wish to
determine here is: what is the subset of $d$ dimensional actions that are
related to each other by the surviving canonical symmetry in the remaining $%
d $ dimensions. As we will see, there are disconnected branches, one for low
spin backgrounds and one for high spin backgrounds. These branches may
correspond to independent theories, or to different phases or limits of the
same theory. Interestingly, string theory seems to offer a possibility of
making a connection to these branches in the zero and infinite tension
limits. Furthermore, we will show that the non-commutative field theory
constructed in \cite{ncsp}, which includes interactions, contains precisely
the same branches in the free limit.

As shown at the end of section 3, a subgroup of the higher-spin symmetries
that keeps the form of $Q_{ij}$ unchanged satisfy
\begin{equation}
\partial W\cdot \varepsilon _{s\geq 1}=0,\quad \mathcal{L}%
_{V_{1}}\varepsilon _{s\geq 1}=0,\quad \partial G_{0}\cdot \varepsilon
_{2}=0.
\end{equation}
We will solve these equations explicitly and identify the unconstrained
remaining symmetry parameters. We will discuss the case for $W=-2\kappa w$
and $V_{1}^{M}=(\kappa ,w,0)$ of the previous subsection. The first equation
becomes
\begin{equation}
w\varepsilon _{s\geq 1}^{M_{1}\cdots M_{s-1}\kappa }=-\kappa \varepsilon
_{s\geq 1}^{M_{1}\cdots M_{s-1}w}  \label{compeps}
\end{equation}
and the second equation becomes
\begin{equation}
\left( \kappa \partial _{\kappa }+w\partial _{w}\right) \varepsilon _{s\geq
1}^{M_{1}\cdots M_{s}}-\sum_{n=1}^{s}\delta _{\kappa }^{M_{n}}\varepsilon
_{s\geq 1}^{M_{1}\cdots M_{n-1}\kappa M_{n+1}\cdots
M_{s}}-\sum_{n=1}^{s}\delta _{w}^{M_{n}}\varepsilon _{s\geq 1}^{M_{1}\cdots
M_{n-1}wM_{n+1}\cdots M_{s}}=0.
\end{equation}
Specializing the indices for independent components and also using (\ref
{compeps}) we get the solution for all components of the higher-spin
symmetry parameters, that obey the subgroup conditions, as
\begin{equation}
\varepsilon _{s\geq 1}^{\overbrace{\kappa \cdots \kappa }^{m}\overbrace{%
w\cdots w}^{n}\mu _{1}\cdots \mu _{s-n-m}}=(-1)^{n}\kappa
^{m}w^{n}\varepsilon _{s,\left( s-n-m\right) }^{\mu _{1}\cdots \mu _{s-n-m}},
\end{equation}
where $\varepsilon _{s,k}^{\mu _{1}\cdots \mu _{k}}\left( x,\frac{w}{\kappa }%
\right) ,$ with $k=0,1,\cdots ,s,$ are arbitrary parameters and independent
of each other. Therefore the form of $\varepsilon \left( X,P\right) $ that
satisfies all the conditions for the remaining symmetry takes the form
\begin{equation}
\varepsilon _{remain}\left( X,P\right) =\sum_{s=0}^{\infty }\sum_{k=0}^{s}
\varepsilon _{s,k}^{\mu _{1}\cdots \mu _{k}}\,\left( \kappa p_{\kappa
}-wp_{w}-\frac{w}{\kappa}B\right) ^{s-k}\,(p_{\mu _{1}}+A_{\mu _{1}})\cdots
(p_{\mu _{k}}+A_{\mu _{k}}).  \label{eremain}
\end{equation}
This identifies $\varepsilon _{s,k}^{\mu _{1}\cdots \mu _{k}}\left( x,\frac{w%
}{\kappa }\right) ,$ with $k=0,1,\cdots ,s,$ as the unconstrained remaining
canonical transformation parameters.

For notational purposes we are going to use the symbol $\varepsilon _{s}^{k}$
for $\varepsilon _{s,k}^{\mu _{1}\cdots \mu _{k}}$ from now on. We will also
indicate the highest-spin fields $g_{s,s}^{\mu _{1}\mu _{2}\cdots \mu _{s}}$
in $d$ dimensions as simply $g_{s}.$ The third condition in (\ref{remain})
gives some extra constraint on $\varepsilon _{2}^{MN}$ which will not be
needed here, so we are going to ignore that condition in the rest of this
discussion.

Let us now consider the gauge $\kappa (\tau )=1$ and $p_{\kappa }(\tau )=0,$
$B=0,$ and the physical sector that satisfies $\tilde{Q}_{11}=\tilde{Q}%
_{12}=0$ (or $w\left( \tau \right) =p_{w}\left( \tau \right) =0$) as
described by the $d$-dimensional holographic picture whose action is (\ref
{actiond}). We discuss the role of the remaining canonical symmetry in this
gauge. The transformation laws for the relevant high-spin fields $g_{s},$
computed from (\ref{deltaQ22}) through $\left\{ \varepsilon _{remain},\tilde{%
Q}_{22}\right\} $, come only from the terms $k=\left( s-1\right) ,s$ in (\ref
{eremain}) since we set $w=p_{w}=p_{\kappa }=0$ and $\kappa =1$ after
performing the differentiation in the Poisson bracket $\left\{ \varepsilon
_{remain},\tilde{Q}_{22}\right\} $. Equivalently, one may obtain the
transformation laws in this gauge by specializing the indices in (\ref
{deltaGs}). The result is
\begin{eqnarray}
\delta g_{s} &=&\left( 2\varepsilon _{1}^{0}g_{s}-\mathcal{L}_{\varepsilon
_{1}^{1}}g_{s}\right)  \nonumber \\
&&+\sum_{n=2}^{s-1}\left( 2\varepsilon _{n}^{n-1}g_{s-n+1}-\varepsilon
_{n}^{n}\cdot \partial g_{s-n+1}+g_{s-n+1}\cdot \partial \varepsilon
_{n}^{n}-\varepsilon _{n}^{n}Fg_{s-n+2}\right)  \nonumber \\
&&-\varepsilon _{s}^{s}Fg_{2}+2(s+1)\varepsilon _{s+1}^{s}u-\varepsilon
_{s+1}^{s+1}\cdot \partial u.
\end{eqnarray}
Each higher-spin field $g_{s}$ is transformed by lower-rank transformation
parameters, $\varepsilon _{n}^{n-1}$ and $\varepsilon _{n}^{n}$ $(n=1,\cdots
,s-1)$, and also by $\varepsilon _{s}^{s}$, $\varepsilon _{s+1}^{s}$ and $%
\varepsilon _{s+1}^{s+1}$. In passing we note that these transformations
inherited from $d+2$ dimensions are somewhat different than those considered
in \cite{segal} although there is some overlap.

If we specialize to $s=2$, we get
\begin{equation}
\delta g_{2}^{\mu \nu }=2\varepsilon _{1}^{0}g_{2}^{\mu \nu }-\mathcal{L}%
_{\varepsilon _{1}^{1}}g_{2}^{\mu \nu }-2\varepsilon _{2}^{\rho (\mu
}F_{\rho \sigma }g_{2}^{\nu )\sigma }+6\varepsilon _{3}^{\mu \nu
}u-3\varepsilon _{3}^{\mu \nu \rho }\partial _{\rho }u.
\end{equation}
Other than the usual general coordinate transformations associated with $%
\varepsilon _{1}^{1}$ and the Weyl dilatations associated with $\varepsilon
_{1}^{0}$, it contains second rank $\varepsilon _{2}^{\rho \mu },\varepsilon
_{3}^{\mu \nu }$ and third rank $\varepsilon _{3}^{\mu \nu \rho }$
transformation parameters. The latter unusual transformations mix the
gravitational field with the gauge field $F_{\rho \sigma }$ and with the
scalar field $\ u.$ Under such transformations, if a field theory with such
local symmetry could exist, one could remove the gravitational field
completely. In fact the same remark applies to all $g_{s}.$ If these could
be true gauge symmetries, all worldline theories would be canonically
transformed to trivial backgrounds. However, there are no known field
theories that realize this local symmetry, and therefore it does not make
sense to interpret them as symmetries in the larger space of $d$ dimensional
worldline theories. This was of concern in \cite{segal}. Fortunately there
is a legitimate resolution by realizing that there are two branches of
worldline theories, one for low spin ($s\leq 2$) and one for high spin ($%
s\geq 2$), that form consistent sets under the transformations as follows.

The first branch is associated with familiar field theories for the low spin
sector including $u,A_{\mu },g_{\mu \nu }.$ The corresponding set of
worldline actions $S(u,A,g_{2})$, in which all background fields $g_{s\geq
3} $ vanish, are transformed into each other under gauge transformations $%
\varepsilon _{0}\left( x\right) ,$ dilatations $\varepsilon _{1}^{0}$ and
general coordinate transformations $\varepsilon _{1}^{1}$. Since $g_{s\geq
3}=0,$ all $\varepsilon _{s\geq 2}$ must be set to zero, and then the low
spin parameters $\varepsilon _{0},\varepsilon _{1}^{0},\varepsilon _{1}^{1}$
form a closed group of local transformations realized on only $u,A_{\mu
},g_{\mu \nu },$ as seen from the transformation laws given above. This
defines a branch of worldline theories for low spins that are connected to
each other by the low spin canonical transformations. This is the usual set
of familiar symmetries and actions.

A second branch of worldline theories exists when the background fields $%
u,A_{\mu }$ vanish. In this high spin branch only $g_{s\geq 2}$ occurs and
therefore, according to the transformations given above they form a basis
for a representation including only the lower rank gauge parameters $%
\varepsilon _{k}^{k-1}$ and $\varepsilon _{k}^{k}$ $(k=1,\cdots ,s-1)$. Then
the transformation rule for the higher-spin fields in $d$ dimensions becomes
\begin{eqnarray}
\delta g_{s\geq 2} &=&\sum_{n=1}^{s-1}\left( 2\varepsilon
_{n}^{n-1}g_{s-n+1}-\varepsilon _{n}^{n}\cdot \partial
g_{s-n+1}+g_{s-n+1}\cdot \partial \varepsilon _{n}^{n}\right)
\label{Wtransf} \\
&=&\left( 2\varepsilon _{1}^{0}g_{s}-\mathcal{L}_{\varepsilon
_{1}^{1}}g_{s}\right) +\left( 2\varepsilon _{2}^{1}g_{s-1}-\varepsilon
_{2}^{2}\cdot \partial g_{s-1}+g_{s-1}\cdot \partial \varepsilon
_{2}^{2}\right) +\cdots  \nonumber \\
&&\cdots +\left( 2\varepsilon _{s-1}^{s-2}g_{2}-\varepsilon
_{s-1}^{s-1}\cdot \partial g_{2}+g_{2}\cdot \partial \varepsilon
_{s-1}^{s-1}\right) .
\end{eqnarray}
We note that the very last term contains $g_{2}^{\mu \nu },$ which is the $d$
dimensional metric that can be used to raise indices
\begin{equation}
\delta g_{s}^{\mu _{1}\mu _{2}\cdots \mu _{s}}=\cdots +\left( 2g_{2}^{(\mu
_{1}\mu _{2}}\varepsilon _{s-1,s-2}^{\mu _{3}\cdots \mu _{s})}-\varepsilon
_{s-1,s-1}^{\mu (\mu _{3}\cdots \mu _{s}}\partial _{\mu }g_{2}^{\mu _{1}\mu
_{2})}+\partial ^{(\mu _{1}}\varepsilon _{s-1,s-1}^{\mu _{2}\cdots \mu
_{s})}\right) .  \label{sgauge}
\end{equation}
The very last term contains the usual derivative term expected in the gauge
transformation laws of a high spin gauge field in $d$ dimensions.

Not all components of the remaining $g_{s}$ can be removed with these gauge
transformations; therefore physical components survive in this high spin
branch. In particular, there is enough remaining freedom to make further
gauge choices such that $g_{s}^{\mu _{1}\mu _{2}\cdots \mu _{s}}$ is double
traceless (i.e. $g_{s\geq 4}^{\mu _{1}\mu _{2}\cdots \mu _{s}}\left(
g_{2}\right) _{\mu _{1}\mu _{2}}\left( g_{2}\right) _{\mu _{3}\mu _{4}}=0$),
as needed for a correct description of high spin fields \cite{fronsdal}. The
high-spin background fields defined in this way belong to a unitary theory.
It is known that with the double traceless condition on $g_{s}$, and the
gauge symmetry generated by traceless $\varepsilon _{s-1,s-1}$ (which is a
subgroup of our case), the correct kinetic terms for high spin fields are
written uniquely in a field theory approach. Thus, the worldline theory
constructed with the double traceless $g_{s}$ makes sense physically. We
would not be allowed to make canonical transformations to further simplify
the worldline theory if we assume that it corresponds to a more complete
theory in which the extra transformations could not be implemented.

Having clarified this point, we may still analyze the fate of the canonical
symmetry left over after the double traceless condition. The remaining gauge
parameters must satisfy the conditions that follow from the double
tracelessness of $\delta g_{s}:$
\begin{equation}
\left( \cdots +2g_{2}^{(\mu _{1}\mu _{2}}\varepsilon _{s-1,s-2}^{\mu
_{3}\cdots \mu _{s})}-\varepsilon _{s-1,s-1}^{\mu (\mu _{3}\cdots \mu
_{s}}\partial _{\mu }g_{2}^{\mu _{1}\mu _{2})}+\partial ^{(\mu
_{1}}\varepsilon _{s-1,s-1}^{\mu _{2}\cdots \mu _{s})}\right) \left(
g_{2}\right) _{\mu _{1}\mu _{2}}\left( g_{2}\right) _{\mu _{3}\mu _{4}}=0.
\end{equation}
If not prevented by some mechanism in a complete theory, this remaining
symmetry is sufficiently strong to make the $g_{s}^{\mu _{1}\mu _{2}\cdots
\mu _{s}}$ not just double traceless, but also traceless. In this case, the
resulting gravity theory would be conformal gravity, which is naively
non-unitary. However, there are ways of curing the problem in a conformal
gravitational field theory setting. One approach is to include
``compensator'' fields that absorb the extra gauge symmetry, thus leaving
behind only the correct amount of symmetry as described in the previous
paragraph. The possibility for such a mechanism appears to be present in the
local Sp$\left( 2,R\right) $ non-commutative field theory formalism of \cite
{ncsp} that includes interactions, and in which $\varepsilon \left(
X,P\right) $ plays the role of gauge symmetry parameters. Indeed, the
background field configurations described so far in the worldline formalism
also emerge in the solution of the non-commutative field equations of this
theory, in the free limit, as described in the following section.

It is also interesting to note that string theory seems to be compatible
with our results. String theory contains two branches of massless states in
two extreme limits, that is, when the string tension vanishes or goes to
infinity, as outlined in the introduction. To better understand this
possible relation to string theory we would have to construct transformation
rules for the extremes of string theory, which are not presently known in
the literature. Hence, the proposed connection to string theory is a
conjecture at this stage. If this connection is verified, it is interesting
to speculate that the high energy, fixed angle, string scattering
amplitudes, computed by Gross and Mende \cite{gross}, may describe the
scattering of a particle in the type of background fields we find in this
paper. Note that an appropriate infinite slope limit $\alpha ^{\prime
}\rightarrow \infty $ can be imitated by the limit $s,t,u\rightarrow \infty $
(at fixed angle) used by Gross and Mende, since $\alpha ^{\prime }$
multiplies these quantities in string amplitudes.

We also find a connection between our transformation rules inherited from $%
d+2$ dimensions, and the transformation rules in $\mathcal{W}$-geometry
analyzed by Hull \cite{Hull} in the special cases of $d=1,2$. The $\mathcal{W%
}$-geometry or generalized Riemannian geometry is defined by a generalized
metric function, on the tangent bundle $TM$ of the target manifold $M,$
which defines the square of the length of a tangent vector $y^{\mu }\in
T_{x}M$ at $x\in M$. The inverse metric is also generalized by introducing a
co-metric function $F(x,y)$ on the cotangent bundle, which is expanded in $y$
as in \cite{Hull}
\begin{equation}
F(x,y)=\sum_{s}\frac{1}{s}\,\,g_{s}^{\mu _{1}\cdots \mu _{s}}(x)\,y_{\mu
_{1}}\cdots y_{\mu _{s}}
\end{equation}
where the coefficients $g_{s}^{\mu _{1}\cdots \mu _{s}}(x)$ are
contravariant tensors on $M$. It is observed in \cite{Hull} that the
coefficients $g_{s}^{\mu _{1}\cdots \mu _{s}}(x)$ in co-metric function can
be associated to higher-spin gauge fields on $M$ only if the co-metric
function is invariant under symplectic diffeomorphism group of the cotangent
bundle of $M$ in $d=1$ and under a subgroup of it in $d=2$. This leads to a
natural set of transformation rules for the gauge fields $g_{s}^{\mu
_{1}\cdots \mu _{s}}(x)$ in dimensions $d=1$ and 2. The transformation rules
that are given in \cite{Hull} for $g_{s}^{\mu _{1}\cdots \mu _{s}}(x)$ in $%
d=1$ and $d=2$ exactly matches the transformation rules (\ref{Wtransf}) that
we found in any dimension by using the 2Tphysics techniques. In the language
of \cite{Hull} the first term in (\ref{Wtransf}) is the $\mathcal{W}$-Weyl
transformation, and the second and the third terms combined are the action
of some subgroup of the symplectic diffeomorhisms of the cotangent bundle of
space-time. We emphasize that our results are valid in any dimension.

\section{Solution of NCFT equation to all orders in $\hbar $}

One may ask the question: which field theory could one write down, such that
its equations of motion, after ignoring field interactions, reproduce the
first quantized version of the physics described by our worldline theory.
That is, we wish to construct the analog of the Klein-Gordon equation
reproducing the first quantization of the relativistic particle. Then in the
form of field theory interactions are included. A non-commutative field
theory (NCFT) formulation of 2Tphysics which addresses and solves this
question is introduced in \cite{ncsp}. The basic ingredient is the local Sp$%
\left( 2\right) $ symmetry, but now in a NC field theoretic setting. The
NCFT equations have a special solution described by the NC field equations (%
\ref{ncft1},\ref{ncft2}). We would like to find all $Q_{ij}\left( X,P\right)
$ that satisfies these equations to all orders of $\hbar $ which appears in
the star products.

It is clear that the classical solution for the background fields discussed
up to now is a solution in the $\hbar \rightarrow 0$ limit, since then the
star commutator reduces to the classical Poisson bracket. However,
surprisingly, by using an appropriate set of coordinates, the classical
solution is also an exact quantum solution. These magical coordinates occur
whenever $W\left( X\right) $ is at the most quadratic in $X^{M}$ and $%
V_{1}^{M}\left( X\right) $ is at the most linear in $X^{M}.$ Thus both of
the cases $W=X^{2}$ and $W=-2\kappa w$ discussed in the previous section
provide exact quantum solutions, and similarly others can be constructed as
well.

To understand this assertion let us examine the transformation rules given
in section 2, but now for general $\hbar $ using the full star product.
Evidently, the classical transformations get modified by all higher orders
in $\hbar $. These are the local Sp$\left( 2\right) $ gauge transformation
rules of the $Q_{ij}$ in the NCFT where $\varepsilon \left( X,P\right) $ is
the local gauge parameter \cite{ncsp}. With these rules we can still map $%
Q_{11}=W\left( X\right) $ as in (\ref{W1}). However, if we proceed in the
same manner as in section 2, since the Poisson bracket would be replaced by
the star commutator everywhere, we are bound to find higher order $\hbar $
corrections in all the expressions. However, consider the star commutator of
$W\left( X\right) $ with any other quantity $\left[ W\left( X\right) ,\cdots %
\right] _{\star }.$ This is a power series containing only odd powers of $%
\hbar .$ If $W\left( X\right) $ is at the most quadratic function of $X^{M},$
the expression contains only the first power of $\hbar $. Hence for
quadratic $W\left( X\right) =X^{2}$ or $W=-2\kappa w$ the star commutator is
effectively replaced by the Poisson bracket, and all expressions involving
such $W\left( X\right) $ produce the same results as the classical analysis.

Similarly, we can argue that, despite the complications of the star product,
we can use the remaining gauge freedom to fix $V_{s\geq 2}=0,$ $V_{0}=0,$ $%
G_{1}=0,$ and $V_{1}^{M}\left( X\right) $ linear in $X^{M}.$ Again, with
linear $V_{1}^{M}\left( X\right) $ all of its star commutators are replaced
by Poisson brackets.

Then, the classical analysis of the background fields, and their
transformation rules, apply intact in the solution of the NCFT field
equations (\ref{ncft1}). The conclusion, again, is that there are two
disconnected branches, one for low spins $s\leq 2$ and one for high spins $%
s\geq 2,$ that seem to have an analog in string theory at the extreme
tension limits.

The NCFT of \cite{ncsp} allow more general field configurations in which the
higher-spin fields interact with each other and with matter to all orders of
$\hbar $ and with higher derivatives, consistently with the gauge
symmetries. In the full theory, the type of field that appears in (\ref
{ncft2}) can play the role of the ``compensators'' alluded to in the
previous section. This would provide an example of an interacting field
theory for higher-spin fields.

\section{Conclusions and remarks}

In this paper it is demonstrated that, in a worldline formalism, all the
usual $d$ dimensional Yang-Mills, gravitational and scalar interactions
experienced by a particle, plus interactions with higher-spin fields, can be
embedded in $d+2$ dimensional 2Tphysics as a natural solution of the
two-time background field equations (\ref{lv1g}), taken in a fixed Sp$\left(
2,R\right) $ gauge. Since 2Tphysics provides many $d$ dimensional
holographic images that appear as different 1T dynamics, a new level of
higher dimensional unification is achieved by the realization that a family
of $d$ dimensional dynamical systems (with background fields) are unified as
a single $d+2$ dimensional theory.

It is also argued that the same perspective is true in field theory provided
we use the NCFT approach to 2Tphysics proposed recently in \cite{ncsp}
which, beyond the worldline theory, provides a coupling of all these gauge
fields to each other and to matter. In the NCFT counterpart the same picture
emerges for a special solution of the NC field equations. Furthermore, the
classical solution that determines the phase space configuration of the
background fields is also a special exact solution of the NCFT equations to
all orders of $\hbar $ when, by using gauge freedom, $W\left( X\right) $ is
chosen as any quadratic function of $X^{M}$ (equivalently, $V_{1}^{M}$ taken
a linear function of $X^{M}).$ In the present paper we gave two
illustrations by taking $W=X^{2}$ and $W=-2w\kappa .$ For non-quadratic
forms of $W\left( X\right) $ there would be higher powers of $\hbar $ in the
solutions of the NCFT equations.

By considering the canonical transformations in phase space in the worldline
formalism (or the gauge symmetry in NCFT formalism) it is argued that a
given solution for a fixed set of background fields can be transformed into
new solutions for other sets of background fields. The physical
interpretation of this larger set of solutions could be very rich, but it is
not investigated in this paper.

The holographic image of the $(d+2)$ dimensional theory, in the massless
particle gauge, makes connections with other formalisms for higher-spin
fields. In particular in one gauge our $d+2$ dimensional approach yields the
$d$ dimensional action discussed in \cite{segal}. As it is shown there, the
first order action (in phase space) is a completed version of an action
originally proposed by de Wit and Freedman \cite{dwitf} in position-velocity
space. The completion consists of including all powers of the velocities
that couple to the higher-spin fields, and their effect in the complete form
of transformation rules. Some problems pointed out in \cite{segal} can be
resolved by three observations: first, there are different branches of
solutions, one for the low spin sector, and one for the high spin sector
starting with spin 2; second, a worldline theory with the correct unitary
high spin fields certainly is permitted as one of the holographic pictures
of the $d+2$ theory; and third, the stronger canonical gauge symmetries that
could lead to non-unitary conformal gravity need not exist in a complete
interacting theory.

Our description of higher-spin fields appears to be consistent in the
worldline formalism, while the non-commutative field theory approach of \cite
{ncsp} provides a field theoretic action for them, with interactions. In
this paper we touched upon this aspect only superficially. This is an old
problem \cite{vasiliev} that deserves further careful study. Furthermore,
our solution may correspond to a self consistent subsectors of string theory
at extreme limits of the tension.

It would also be very interesting to further study the holographic aspects
of the 2Tphysics theory.

\section{Acknowledgments}

I.B. would like to thank Edward Witten, Misha Vasiliev and Djordje Minic for
discussions. This research was in part supported by the US Department of
Energy under grant number DE-FG03-84ER40168, and by the CIT-USC Center of
Theoretical Physics.

\end{document}